\newif\ifdraft
\drafttrue

\newif\ifarxiv
\newif\ifone

\onefalse

\arxivtrue

\ifone 
\documentclass[journal,12pt,onecolumn,draftclsnofoot,]{IEEEtran}
\else 
\documentclass[conference]{IEEEtran}
\IEEEoverridecommandlockouts
\fi

\ifarxiv
\else
\fi

\usepackage{acro}
\acsetup{first-style=all}

\DeclareAcronym{URLLC}{
  short = URLLC ,
  long  = Ultra-Reliable Low Latency Communication ,
  class = abbrev
}

\DeclareAcronym{IR}{
    short = {IR} ,
    long  = Incremental Redundancy ,
    class = abbrev
}

\DeclareAcronym{EEG}{
    short = {EEG} ,
    long  = Energy Efficiency Gain ,
    class = abbrev
}

\DeclareAcronym{DMRS}{
  short = DMRS ,
  long  = Demodulation Reference Signal,
  class = abbrev
}

\DeclareAcronym{WI}{
  short = WI ,
  long  = Work Item,
  class = abbrev
}

\DeclareAcronym{QAM}{
  short = QAM ,
  long  = Quadrature Amplitude Modulation ,
  class = abbrev
}

\DeclareAcronym{OFDM}{
  short = OFDM ,
  long  = Orthogonal Frequency Division Multiplexing ,
  class = abbrev
}

\DeclareAcronym{OFDMA}{
  short = OFDMA ,
  long  = Orthogonal Frequency Division Multiplexing Access ,
  class = abbrev
}

\DeclareAcronym{E2E}{
  short = E2E ,
  long  = End-to-End ,
  class = abbrev
}

\DeclareAcronym{DL}{
  short = DL ,
  long  = DownLink ,
  class = abbrev
}

\DeclareAcronym{PDCP}{
  short = PDCP ,
  long  = Packet Data Convergence Protocol ,
  class = abbrev
}

\DeclareAcronym{prHARQ}{
  short = prHARQ ,
  long  = proactive HARQ with prediction ,
  class = abbrev
}

\DeclareAcronym{eprHARQ}{
  short = eprHARQ ,
  long  = early proactive HARQ with prediction ,
  class = abbrev
}

\DeclareAcronym{paHARQ}{
  short = paHARQ ,
  long  = proactive HARQ ,
  class = abbrev
}

\DeclareAcronym{reHARQ}{
  short = reHARQ ,
  long  = reactive HARQ ,
  class = abbrev
}

\DeclareAcronym{GF}{
  short = GF ,
  long  = Grant-Free ,
  class = abbrev
}

\DeclareAcronym{RB}{
  short = RB ,
  long  = Resource Block ,
  class = abbrev
}

\DeclareAcronym{RAN}{
  short = RAN ,
  long  = Radio Access Network ,
  class = abbrev
}

\DeclareAcronym{MCS}{
  short = MCS ,
  long  = Modulation and Coding Scheme ,
  class = abbrev
}

\DeclareAcronym{CFI}{
  short = CFI ,
  long  = Control Format Indicator ,
  class = abbrev
}

\DeclareAcronym{UL}{
  short = UL ,
  long  = UpLink ,
  class = abbrev
}

\DeclareAcronym{SR}{
  short = SR ,
  long  = Scheduling Request ,
  class = abbrev
}

\DeclareAcronym{BER}{
  short = BER ,
  long  = Bit Error Rate ,
  class = abbrev
}

\DeclareAcronym{BLER}{
  short = BLER ,
  long  = Block Error Rate ,
  class = abbrev
}

\DeclareAcronym{CRC}{
  short = CRC ,
  long  = Cyclic Redundancy Check ,
  class = abbrev
}

\DeclareAcronym{CB}{
  short = CB ,
  long  = Code Block ,
  class = abbrev
}

\DeclareAcronym{CBG}{
  short = CBG ,
  long  = Code Block Group ,
  class = abbrev
}

\DeclareAcronym{R-CBG}{
  short = R-CBG ,
  long  = Reduced Code Block Group ,
  class = abbrev
}

\DeclareAcronym{AR-CBG}{
  short = AR-CBG ,
  long  = Adaptive Reduced Code Block Group ,
  class = abbrev
}

\DeclareAcronym{TB}{
  short = TB ,
  long  = Transport Block ,
  class = abbrev
}

\DeclareAcronym{BG2}{
  short = BG2 ,
  long  = Base Graph 2 ,
  class = abbrev
}

\DeclareAcronym{BG1}{
  short = BG1 ,
  long  = Base Graph 1 ,
  class = abbrev
}

\DeclareAcronym{SNR}{
  short = SNR ,
  long  = Signal-to-Noise Ratio ,
  class = abbrev
}

\DeclareAcronym{CCE}{
  short = CCE ,
  long  = Control Channel Element ,
  class = abbrev
}

\DeclareAcronym{PDCCH}{
  short = PDCCH ,
  long  = Physical Downlink Control Channel ,
  class = abbrev
}

\DeclareAcronym{PUCCH}{
  short = PUCCH ,
  long  = Physical Uplink Control Channel ,
  class = abbrev
}

\DeclareAcronym{LTE}{
  short = LTE ,
  long  = Long Term Evolution ,
  class = abbrev
}

\DeclareAcronym{NGMN}{
  short = NGMN ,
  long  = Next Generation Mobile Networks ,
  class = abbrev
}

\DeclareAcronym{RNTI}{
  short = RNTI ,
  long  = Radio Network Temporary Identifier ,
  class = abbrev
}

\DeclareAcronym{3GPP}{
  short = 3GPP ,
  long  = 3rd Generation Partnership Project ,
  class = abbrev
}

\DeclareAcronym{HRLLC}{
  short = HRLLC ,
  long  = High-Reliable Low Latency Communication ,
  class = abbrev
}

\DeclareAcronym{TTI}{
  short = TTI ,
  long  = Transmission Time Interval ,
  class = abbrev
}

\DeclareAcronym{sTTI}{
  short = sTTI ,
  long  = short Transmission Time Interval ,
  class = abbrev
}

\DeclareAcronym{RTT}{
  short = RTT ,
  long  = Round Trip Time ,
  class = abbrev
}

\DeclareAcronym{LDPC}{
  short = LDPC ,
  long  = Low-Density Parity-Check ,
  class = abbrev
}

\DeclareAcronym{UE}{
  short = UE ,
  long  = User Equipment ,
  class = abbrev
}

\DeclareAcronym{TI}{
    short = TI ,
    long  = Tactile Internet ,
    class = abbrev
}

\DeclareAcronym{BS}{
  short = BS ,
  long  = Base Station ,
  class = abbrev
}

\DeclareAcronym{FPR}{
  short = FPR ,
  long  = False-Positive Rate ,
  class = abbrev
}

\DeclareAcronym{FNR}{
  short = FNR ,
  long  = False-Negative Rate ,
  class = abbrev
}

\DeclareAcronym{DCI}{
  short = DCI ,
  long  = Downlink Control Information ,
  class = abbrev
}

\DeclareAcronym{HARQ}{
  short = HARQ ,
  long  = Hybrid Automatic Repeat reQuest ,
  class = abbrev
}

\DeclareAcronym{IIOT}{
  short = IIOT ,
  long  = Industrial Internet of Things ,
  class = abbrev
}

\DeclareAcronym{RV}{
  short = RV ,
  long  = Redundancy Version ,
  class = abbrev
}

\DeclareAcronym{ACK}{
  short = ACK ,
  long  = ACKnowledgment ,
  class = abbrev
}

\DeclareAcronym{NACK}{
  short = NACK ,
  long  = Non-ACKnowledgment ,
  class = abbrev
}

\DeclareAcronym{CG}{
  short = CG ,
  long  = Configured Grant ,
  class = abbrev
}

\DeclareAcronym{E-HARQ}{
  short = E-HARQ ,
  long  = Early HARQ ,
  class = abbrev
}

\DeclareAcronym{P-HARQ}{
  short = P-HARQ ,
  long  = Predictive incremental-redundancy rateless HARQ ,
  class = abbrev
}

\DeclareAcronym{R-HARQ}{
  short = R-HARQ ,
  long  = Rateless incremental-redundancy HARQ ,
  class = abbrev
}

\DeclareAcronym{C-HARQ}{
  short = C-HARQ ,
  long  = Conventional incremental-redundancy HARQ ,
  class = abbrev
}

\DeclareAcronym{mMTC}{
  short = mMTC ,
  long  = massive Machine Type Communications ,
  class = abbrev
}

\DeclareAcronym{5G}{
  short = 5G ,
  long  = Fifth Generation ,
  class = abbrev
}

\DeclareAcronym{SPS}{
  short = SPS ,
  long  = Semi-Persistent Scheduling ,
  class = abbrev
}

\DeclareAcronym{PI}{
  short = PI ,
  long  = Pre-emption Indication ,
  class = abbrev
}

\DeclareAcronym{NR}{
  short = NR ,
  long  = New Radio ,
  class = abbrev
}

\DeclareAcronym{eMBB}{
  short = eMBB ,
  long  = enhanced Mobile BroadBand ,
  class = abbrev
}

\DeclareAcronym{LLR}{
  short = LLR ,
  long  = Log-Likelihood Ratio ,
  class = abbrev
}

\DeclareAcronym{VNR}{
  short = VNR ,
  long  = Variable Node Reliability ,
  class = abbrev
}

\DeclareAcronym{BSC}{
  short = BSC ,
  long  = Binary Symmetric Channel ,
  class = abbrev
}

\DeclareAcronym{angelsperarea}{
  short = \ensuremath{a} ,
  long  = The number of angels per unit area ,
  sort  = a ,
  class = nomencl
}
\DeclareAcronym{numofangels}{
  short = \ensuremath{N} ,
  long  = The number of angels per needle point ,
  sort  = N ,
  class = nomencl
}
\DeclareAcronym{areaofneedle}{
  short = \ensuremath{A} ,
  long  = The area of the needle point ,
  sort  = A ,
  class = nomencl
}
\usepackage{units}
\usepackage{cite}
\usepackage{amsmath,amssymb,amsfonts,amsthm}
\usepackage{bbm}
\usepackage{graphicx}
\usepackage{subfigure}
\usepackage{color}
\usepackage{comment}
\usepackage{multirow}
\usepackage{hhline}
\usepackage[normalem]{ulem} 
\newcommand{\stkout}[1]{\ifmmode\text{\sout{\ensuremath{#1}}}\else\sout{#1}\fi}
\ifdraft
\newcommand{\added}[1]{\textcolor{blue}{#1}}
\newcommand{\deleted}[1]{\textcolor{blue}{\stkout{#1}}}

\newcommand{\deletedfloat}[1]{\textcolor{blue}{#1}}
\newcommand{\cmm}[1]{\textcolor{red}{#1}}
\else
\newcommand{\added}[1]{#1}
\newcommand{\deleted}[1]{}

\newcommand{\deletedfloat}[1]{}
\newcommand{\cmm}[1]{}
\fi

\graphicspath{{./figures/}}

\def\BibTeX{{\rm B\kern-.05em{\sc i\kern-.025em b}\kern-.08em
    T\kern-.1667em\lower.7ex\hbox{E}\kern-.125emX}}

\begin{document}
\title{Feedback Prediction for Proactive HARQ in the Context of Industrial Internet of Things}
\author{\IEEEauthorblockN{Bar{\i}{\c s} G{\" o}ktepe}
\IEEEauthorblockA{\textit{Fraunhofer Heinrich Hertz Institute}\\
Berlin, Germany \\
baris.goektepe@hhi.fraunhofer.de}
\and
\IEEEauthorblockN{Tatiana Rykova}
\IEEEauthorblockA{\textit{Fraunhofer Heinrich Hertz Institute}\\
Berlin, Germany \\
tatiana.rykova@hhi.fraunhofer.de}
\and
\IEEEauthorblockN{Thomas Fehrenbach}
\IEEEauthorblockA{\textit{Fraunhofer Heinrich Hertz Institute}\\
Berlin, Germany \\
thomas.fehrenbach@hhi.fraunhofer.de}
\and
\IEEEauthorblockN{Thomas Schierl}
\IEEEauthorblockA{\textit{Fraunhofer Heinrich Hertz Institute}\\
Berlin, Germany \\
thomas.schierl@hhi.fraunhofer.de}
\and
\IEEEauthorblockN{Cornelius Hellge}
\IEEEauthorblockA{\textit{Fraunhofer Heinrich Hertz Institute}\\
Berlin, Germany \\
cornelius.hellge@hhi.fraunhofer.de}
}

\author{\IEEEauthorblockN{Bar{\i}{\c s} G{\" o}ktepe\IEEEauthorrefmark{1},
Tatiana Rykova\IEEEauthorrefmark{1},
Thomas Fehrenbach\IEEEauthorrefmark{1}, 
Thomas Schierl\IEEEauthorrefmark{1} and
Cornelius Hellge\IEEEauthorrefmark{1}}
\IEEEauthorblockA{\IEEEauthorrefmark{1}Video Coding \& Analytics Department\\
Fraunhofer Heinrich Hertz Institute,
Berlin, Germany\\ Email: first.last@hhi.fraunhofer.de}}

\IEEEpubid{\begin{minipage}{\textwidth}\ \\[12pt]
 978-1-7281-8298-8/20/\$31.00 \copyright 2020 IEEE\\ 
 2020 IEEE GLOBECOM, Taipei, Taiwan
\end{minipage}} 

\maketitle

\begin{abstract}
In this work, we investigate proactive Hybrid Automatic Repeat reQuest (HARQ) using link-level simulations for multiple packet sizes, modulation orders, BLock Error Rate (BLER) targets and two delay budgets of 1~ms and 2~ms, in the context of Industrial Internet of Things (IIOT) applications. In particular, we propose an enhanced proactive HARQ protocol using a feedback prediction mechanism. We show that the enhanced protocol achieves a significant gain over the classical proactive HARQ in terms of energy efficiency for almost all evaluated BLER targets at least for sufficiently large feedback delays. Furthermore, we demonstrate that the proposed protocol clearly outperforms the classical proactive HARQ in all scenarios when taking a processing delay reduction due to the less complex prediction approach into account, achieving an energy efficiency gain in the range of 11\% up to 15\% for very stringent latency budgets of 1~ms at $10^{-2}$ BLER and from 4\% up to 7.5\% for less stringent latency budgets of 2~ms at $10^{-3}$ BLER. Furthermore, we show that power-constrained proactive HARQ with prediction even outperforms unconstrained reactive HARQ for sufficiently large feedback delays.
\end{abstract}

\section{Introduction}
\subsection{Background and Motivation}
Since the completion of Rel.~15 and Rel.~16 \ac{5G} specifications by the \ac{3GPP}, the focus has been shifted from rather generic solutions to cater for the main use cases to more sophisticated solutions targeting the emerging use cases. \ac{TI} is one of the focus areas of the \ac{3GPP}. Especially machine-to-machine type communication, puts demanding requirements on the latency and reliability while keeping the device complexity and power consumption low. Hence, study and work items have been approved to evaluate reduced complexity devices in the context of \ac{IIOT} services\cite{iiot_wid,nrlight_sid}. The frequently mentioned \ac{URLLC} use case with a packet error rate smaller than $10^{-5}$ and 1~ms end-to-end latency \cite{5g_paper} is one of the main goals of the further enhancements. However, as previously mentioned, use cases with more stringent latency targets have also turned towards low-complexity and battery powered devices, e.g. safety related sensors \cite{nrlight_sid}. Here, complexity constraints, such as energy or bandwidth limitation, should also be ensured given the demanding \ac{URLLC} targets. In this work, we focused mainly on the latency and energy constraints since reliability can also be partly moved to higher layer protocols exploiting link diversity, such as \ac{PDCP} duplication \cite{urllc_tr}.\\
Lately, \ac{HARQ} has been extensively investigated in the context of \ac{URLLC} in \cite{urllc_tr}, which resulted in a follow-up work item specification. \ac{HARQ} is a physical layer retransmission mechanism, which enhances spectral efficiency significantly especially in a high reliability regime. However, the major drawback of the so-called reactive \ac{HARQ}, which is one of the fundamental mechanisms of \ac{LTE} and \ac{5G}, is an introduced latency due to the \ac{HARQ} \ac{RTT}. It has to be noted that the reactive \ac{IR} \ac{HARQ} approaches the ergodic channel capacity of Rayleigh block fading channels\cite{5439316} and shows a significantly higher energy efficiency compared to other \ac{HARQ} combination schemes, such as chase-combining \cite{computers7040048}. According to this scheme the transmitter splits the codeword, which is generated by the so-called mother code, into subcodewords also designated as \acp{RV} in the \ac{3GPP} context. These \acp{RV} are transmitted one by one based on the receiver's feedback, where ACK indicates a successful decode and NACK \added{-} an unsuccessful one. However, the \ac{HARQ} \ac{RTT} is a major bottleneck of the reactive \ac{HARQ}\cite{gf_harq_oulu}, and can be reduced by shortening \ac{TTI} to one \ac{OFDM} symbol\cite{gf_harq_oulu}. Nevertheless, this imposes higher requirements on both the  receiver's instantaneous processing bandwidth and the transmitter's power constraint. Furthermore, even performing the transmissions on small multiples of \ac{OFDM} symbols can reduce the latency only to a certain limit defined by the \ac{HARQ} \ac{RTT}. Hence, new \ac{HARQ} solutions that enable low-latency and reduce the instantaneous bandwidth for low complexity devices are of great importance for investigation.\\ 
In Rel.~16 a new \ac{HARQ} mechanism designated as proactive \ac{HARQ} was introduced for \ac{UL} \ac{GF} communication. In compliance with this scheme, the transmitter determines a number of \acp{RV} that
is required to achieve the reliability target. This number of \acp{RV} is transmitted autonomously unless a positive ACK is received before\cite{gf_harq_oulu, gf_harq_aalborg}. This scheme solves the \ac{HARQ} \ac{RTT} latency issue and provides high reliability at very low latency constraints at the expense of spectral efficiency loss \cite{liu2020analyzing}. However, it may lead to the transmission of unnecessary \acp{RV} due to the feedback delay, which is comprised of the delay components for processing, queuing in the upper layers, propagation delays and feedback transmission time.\\
Strategies for reducing the feedback delay using prediction mechanisms have been broadly studied in literature. Different \ac{HARQ} feedback prediction methods based on estimating the channel state are proposed in \cite{caire_fast_harq, zhou_ieeetran, phdthesis_csi_harq, snr_prediction}. Furthermore, authors in \cite{caire_fast_harq} investigate a mixture of proactive and reactive \ac{HARQ} protocols to reduce the expected latency. In particular, the impact of prediction errors is studied in \cite{zhou_ieeetran}. In \cite{early_harq_schemes2,early_harq_schemes}, the authors use a \ac{BER} estimate based on \acp{LLR} to predict the decoding outcome ahead of the actual decoding. Authors in \cite{caire_early_decoding} analyze an approach where an unnecessary decoding of \ac{LDPC} codes is avoided in case of predicted packet loss and thus, the feedback delay is reduced. The machine learning techniques that predict the decoding outcome ahead of full reception of the codeword are studied in \cite{journal_eharq_paper}. In \cite{prediction_cran1} and \cite{prediction_cran2}, authors put the prediction methods into the context of cloud-\acp{RAN} and show the benefits of early feedback in scenarios with a non-ideal backhaul.
\subsection{Contributions and Organization}
The contributions of this paper are summarized as follows:
\begin{itemize}
    \item We put forward the feedback prediction approach using a logistic regression on \ac{LDPC} subcodes \cite{journal_eharq_paper} proposing an enhanced proactive \ac{HARQ} protocol, and evaluate its performance on TDL-C fading channels using link-level simulations for different scenarios, i.e. target \acp{BLER}, feedback delays, \ac{TB} sizes, and delay budgets.
    \item In the second part, we evaluate further benefits of using a feedback prediction in the context of proactive \ac{HARQ} considering a reduction of the feedback delay compared to the full decoding. We show that reducing the feedback delay by employing feedback prediction provides a significant benefit especially for low \acp{BLER} of up to 14\% and up to 8\% for 1~ms and 2~ms delay budgets, respectively.
    \item Furthermore, we show that the proposed proactive \ac{HARQ} scheme while catering to power constraints, also outperforms reactive \ac{HARQ} without any power constraint at the same $E_b/N_0$ ratio for sufficiently high feedback delays relative to the total latency budget, especially in the low \ac{BLER} regime.
\end{itemize}
The paper is organized as follows. Section~\ref{sec:sys_model} presents a detailed setup of different \ac{HARQ} approaches and provides the assumptions for the link-level simulations. Furthermore, specific optimizations for different \ac{HARQ} approaches are presented. In Section~\ref{sec:eval_method} the evaluation methodology for the purpose of performance comparison is explained. In Section~\ref{sec:results}, the results of the simulations are presented. In the first part, proactive \ac{HARQ} with and without prediction is compared under the assumption that the feedback delay parameter is fixed. In the second part, the performance of proactive \ac{HARQ} is analyzed with and without prediction, assuming a shortened feedback delay for the prediction-based scheme, whereas in the last part proactive \ac{HARQ} is compared to reactive \ac{HARQ}.
\section{System Model}
\label{sec:sys_model}
\begin{table}[t]
\centering
\caption{Link-level simulation assumptions for training and test set generation.}
\label{tab:lls_assump}
\begin{tabular}{l|l}
Number of \acp{TB}&1.7M (0.85M train, 0.85M test)\\
\hline
\ac{TB} size in bits ($N_\mathrm{Bits}$) & 360, 500, 800, 1000 \\
\hline
Transmission duration \ac{RV} ($\delta_\mathrm{t}$)&1 \ac{OFDM} symbol \\
\hline
Delay budgets ($\delta_\mathrm{bud}$)&14~$\delta_\mathrm{t}$, i.e. 1~ms,\\
&28~$\delta_\mathrm{t}$, i.e. 2~ms\\
\hline
Feedback delays ($\delta_\mathrm{fb}$)&2 - 5 for $\delta_\mathrm{bud} = 14$,\\
&5 - 12 for $\delta_\mathrm{bud} = 28$\\
\hline
Target \acp{BLER} ($\epsilon_\mathrm{target}$)&$10^{-1}$, $5\cdot 10^{-2}$, $10^{-2}$ ($\delta_\mathrm{bud}=14$)\\
&$10^{-1}$, $10^{-2}$, $10^{-3}$ ($\delta_\mathrm{bud}=28$)\\
\hline
Transmission bandwidth&1.08~MHz (6 RBs)\\
\hline
Channel Code&Rate-1/5 LDPC (see \cite{5g_channel_coding_spec})\\
\hline
Modulation order and algorithm&4-QAM, 16-QAM, 64-QAM,\\
&Approximated LLR\\
\hline
Power allocation&Constant $E_b/N_0$\\
\hline
Waveform&3GPP OFDM,\\
&normal cyclic-prefix,\\
&15~kHz subcarrier spacing\\
\hline
Channel type&1~Tx 1~Rx, TDL-C 100~ns,\\
&2.9~GHz, 3.0~km/h\\
\hline
Equalizer&Frequency domain MMSE\\
\hline
Decoder type&Min-Sum (50 iterations)\\
\hline
Prediction iterations&5
\end{tabular}
\label{tab:SIM}
\end{table}
\raggedbottom

Table~\ref{tab:SIM} summarizes the link-level parameters which have been used for the simulations. We analyzed two different latency constraints, a short delay budget of one slot, i.e. 14 \ac{OFDM} symbols, and a long delay budget of two slots, i.e. 28 \ac{OFDM} symbols. However, these delay budgets can be arbitrarily scaled. Under these constraints, we evaluated three different \ac{HARQ} approaches, i.e. \ac{paHARQ}, \ac{prHARQ} and \ac{reHARQ} for multiple \ac{TB} sizes, a variety of target \ac{BLER} $\epsilon_\mathrm{target}$, and different assumptions on the feedback delay $\delta_\mathrm{fb}$. The \acp{SNR} for the specific scenarios have been chosen such that the target \ac{BLER} is achievable. It is important to note that the feedback delay is comprised of multiple delay components, such as the propagation delay, the processing delay, i.e. decoding, and the feedback delay. The prediction is able to reduce the processing delay, by performing only 5 instead of 50 decoding iterations \cite{journal_eharq_paper}. To maintain the full picture of the proposed approach, we evaluated the performance of the prediction-based scheme under two scenarios: with typical processing delay and a shortened one. In Section~\ref{sec:pharq-rharq}, we assumed that the feedback delay stays the same, designated as \ac{prHARQ}, to show that there is an advantage for a certain parameter range even under this setup. In Section~\ref{sec:early_feedback}, we compared the performance of the prediction-based approach assuming that the shorter processing delay reduces the feedback delay by one \ac{OFDM} symbol, designated as \ac{eprHARQ}.

\subsection{Evaluation Methodology}
\label{sec:eval_method}
In addition to achieving the reliability and the latency targets which are mandatory requirements, the performance of the \ac{HARQ} schemes can be compared in terms of energy efficiency, which is critical for battery powered devices. Based on the renewal-reward theorem \cite{ross1996stochastic}, the energy efficiency as the number of bits that can be transmitted with a given amount of energy is expressed as:
\begin{equation}
\eta = \frac{\mathbb{E}[\mathcal{R}]}{\mathbb{E}[\psi]},
\end{equation}
where $\mathbb{E}[\mathcal{R}]$ is an expected reward and $\mathbb{E}[\psi]$ is an expected consumed energy, where $\mathbb{E}[\psi] := \mathbb{E}[T_\mathrm{HARQ} \cdot P_\mathrm{RV}]$ with $T_\mathrm{HARQ}$ is a number of required transmissions, as defined for the different \ac{HARQ} schemes in the further in the subsections, and $P_\mathrm{RV}$ is a consumed energy for an \ac{RV} which is constant for the proactive \ac{HARQ} schemes. Furthermore, the reward is $\mathcal{R} := 0$ in case the transmission failed within the latency budget and $\mathcal{R} := N_\mathrm{Bits}$ in case the transmission was successful. Hence, the expected reward is given as $\mathbb{E}[\mathcal{R}] := (1 - \epsilon) N_\mathrm{Bits}$, where $\epsilon$ is an associated total error probability.\\
In order to compare different HARQ schemes, we evaluated the \ac{EEG} defined as:
\begin{equation}
    \Theta_\mathrm{H1,H2} := \frac{\eta_\mathrm{H1} - \eta_\mathrm{H2}}{\eta_\mathrm{H1}},
\end{equation}
where $\eta_\mathrm{H1}$ and $\eta_\mathrm{H2}$ are the expected energy efficiency values of the two selected \ac{HARQ} approaches.

\subsection{Reactive HARQ}
\begin{figure}
  \centering
  \includegraphics[width=0.35\textwidth]%
    {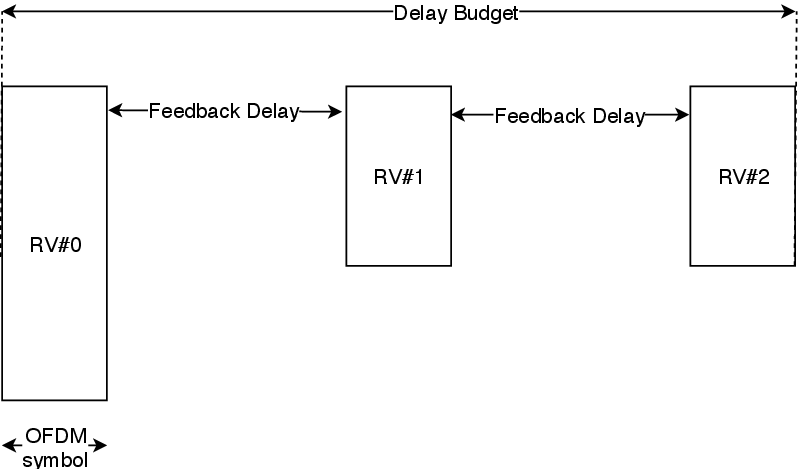}
    \caption{Schematic representation of \ac{reHARQ} without any power or bandwidth constraint.}
    \label{fig:c_harq_schematic}
\end{figure}
The \ac{reHARQ} approach is based on \ac{IR} \ac{HARQ} \cite{gf_harq_oulu}, and as was mentioned before is not able to achieve the required \ac{BLER} target within the given power limitation at the evaluated \ac{SNR} values. However, since it is a widely used state-of-the-art mechanism in \ac{5G}, a comparison is justified although the constraints are not the same. As depicted in Fig.~\ref{fig:c_harq_schematic}, each of the data re-/transmissions can be reallocated to a transmission, which is as short as a single \ac{OFDM} symbol, to reduce the delay as much as possible. Furthermore, the distribution of redundancy over the different re-/transmissions can be optimized such that the expected number of overall transmissions is minimized. The expected number of transmissions, where a transmission is equivalent to an \ac{RV} for the proactive approaches to ensure comparability, is defined as:
\begin{equation}
    \mathbb{E}[T_\mathrm{reHARQ}] = \sum_{i=0}^{n_\mathrm{max}} \epsilon_i T_i,
\end{equation}
where $n_\mathrm{max}$ is a maximum number of \ac{HARQ} re-transmissions within the latency budget and $\epsilon_i$\added{,} and $T_i$ are an associated error probability of a single transmission and a number of \ac{paHARQ}-equivalent \acp{RV}, respectively.

\subsection{Proactive HARQ}
\begin{figure}
  \centering
  \includegraphics[width=0.35\textwidth]%
    {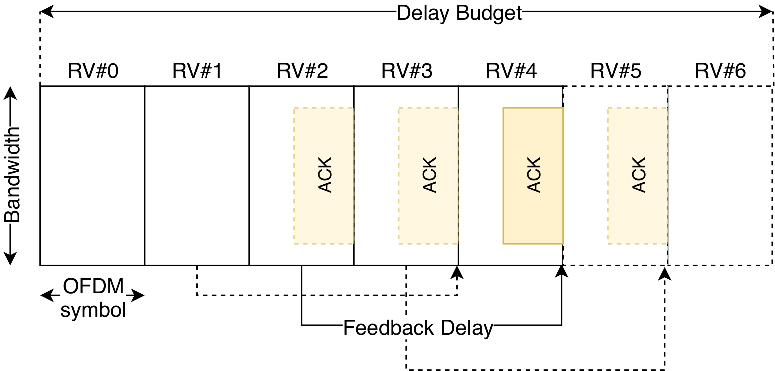}
\caption{Schematic representation of \ac{paHARQ} with bandwidth limitation.}
\label{fig:proactive_harq}
\end{figure}
In Rel.~16 \ac{URLLC}, \ac{3GPP} adopted different \ac{HARQ} approaches including proactive \ac{HARQ} to enable \ac{HARQ} for \acp{CG} in latency constrained scenarios \cite{gf_harq_oulu}. 
The proactive \ac{HARQ} approach first determines a number of \acp{RV} to achieve the target error rate. As depicted in Fig.~\ref{fig:proactive_harq}, these \acp{RV} are transmitted consecutively up to a reception of an ACK which terminates the transmission process. This approach ensures the reception of the packet within the delay budget. However, the efficiency is determined by the feedback delay, also designated as the \ac{HARQ} \ac{RTT}. Since the receiver has to process the received signal stream to generate the feedback, at the time moment when the ACK message reaches the transmitter, multiple \acp{RV} have already been sent. Thus, despite overcoming the \ac{HARQ}-specific issue of the feedback delay in latency constrained scenarios, the gained latency comes from trading off the spectral efficiency due to unnecessary retransmissions. Hence, the expected number of transmissions is designated as:
\begin{align}
    \mathbb{E}[T_\mathrm{paHARQ}] = \sum_{i=1}^{n-\delta_\mathrm{fb}} P_i (i + \delta_\mathrm{fb}) ,
\end{align}
where $n$ is a maximum number of transmissions, $\delta_\mathrm{fb}$ is the feedback delay, and $P_i := \left(\prod_{k=1}^{i-1}\epsilon_k\right) (1 - \epsilon_i) $ is the probability that the packet is decodable at the $i$-th \ac{RV} with $\epsilon_n := 0$.

\subsection{Novel Proactive HARQ with Prediction}
The proactive \ac{HARQ} with prediction uses the same setup, as depicted in Fig.~\ref{fig:proactive_harq}. The major difference is that the full decoding of an \ac{RV} is replaced firstly by a small number of decoding iterations on the received subcode, and secondly, a logistic regression predictor that uses Variable Node Reliabilities (VNRs) as an input vector \cite{ldpc_subcodes,journal_eharq_paper}. Previous work has showed that a logistic regression achieves the best performance among a variety of predictors \cite{journal_eharq_paper}. Due to the paper's length limitation we leave the basic definitions, such as subcode, false negative/positive probabilities, out of scope of this paper. Please refer to \cite{ldpc_subcodes} for more details.
\begin{figure}
  \centering
  \includegraphics[width=0.35\textwidth]%
    {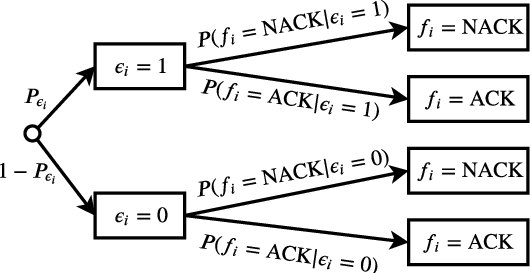}
\caption{Probabilistic model of \ac{HARQ} feedback prediction.}
\label{fig:prob_model}
\end{figure}

Fig.~\ref{fig:prob_model} shows a probabilistic model, which has been used to evaluate the performance of the \ac{prHARQ} approach. In this model the error probabilities have been obtained by Monte-Carlo link-level simulations, and the false-positive and false-negative prediction probabilities were found by firstly training the logistic regression on half of the simulation data and then deriving them based on the remaining data set.

\subsubsection{Parameter selection for Predictive IR HARQ}
False predictions, false-positive as well as false-negative, impact the performance of the \ac{prHARQ} scheme. False-positive mispredictions increase the total \ac{BLER}, whereas false-negative mispredictions mainly cause unnecessary retransmissions, and hence degrade the energy efficiency. In order to optimize the \ac{HARQ} operation the following optimization problem has to be solved:
\begin{equation}
\label{eq:prHARQ_opt}
    \begin{aligned}
        &\underset{P_{\mathrm{fp}}^n}{\text{minimize}}&&\mathbb{E}[T_\mathrm{prHARQ}(P_{\mathrm{fp}}^n)]\\
        &\text{subject to} &&\epsilon(P_{\mathrm{fp}}^n) \leq \epsilon_{\mathrm{target}},
    \end{aligned}
\end{equation}
where $P_{\mathrm{fp}}^n := (P_{\mathrm{fp}_1},...,P_{\mathrm{fp}_i},...,P_{\mathrm{fp}_n})$ with $P_{\mathrm{fp}_i} := P(f_i=\mathrm{ACK}|\epsilon_i=1)$ are the false-positive prediction error probabilities, which are used to compute the false-negative prediction error probabilities: $P_{\mathrm{fn}_i} := P(f_i=\mathrm{NACK}|\epsilon_i=0)=f(P_{\mathrm{fp}_i})$.
Furthermore, $T(P_{\mathrm{fp}}^n) \in \{T_\mathrm{min}+\delta_\mathrm{fb},T_\mathrm{min}+\delta_\mathrm{fb}+1,...,T_\mathrm{max}\}$ is a random variable representing the number of required transmissions. For the purpose of simplification, let $K \in \{0,...,k_\mathrm{max}\}$ be a dependent random variable with $T(P_{\mathrm{fp}}^n)=T_\mathrm{min}+K+\delta_\mathrm{fb}$ and $k_{\mathrm{max}} := T_{\mathrm{max}}-T_{\mathrm{min}}-\delta_\mathrm{fb}$. Hence, the probability distribution of $T(P_{\mathrm{fp}}^n)$ is given as follows:
\begin{align}
\mathbb{P}[&T(P_{\mathrm{fp}}^n) = t]:= \mathbb{P}[K = k] = P_{1}(k),
\end{align}
where $P_{i}(k)$ with $i \in \{1, ..., k_\mathrm{max}+1\}$ is $P_{k+1}(k) := 1$ for $i=k+1$ and $k=k_\mathrm{max}$, and otherwise is defined recursively as:
\begin{align}
P_i(k) &:=
\begin{cases}
P_{\epsilon_i}(1-P_{\mathrm{fp}_i})P_{i+1},&\text{if }i<k,\\
P_{\epsilon_i}(1-P_{\mathrm{fp}_i})P_{i+1}+(1-P_{\epsilon_i})P_{\mathrm{fn}_i},&\text{if }i=k,\\
P_{\epsilon_i}(1-P_{\mathrm{fp}_i})+(1-P_{\epsilon_i})P_{\mathrm{fn}_i},&\text{if }i=k+1.\\
\end{cases}
\end{align}
Furthermore, the total \ac{BLER} $\epsilon(P_{\mathrm{fp}}^n)$ is defined as follows:
\begin{align}
    \epsilon(P_{\mathrm{fp}}^n)&:= \left(\prod_{i=1}^{\delta_\mathrm{fb}}P_{\epsilon_i}\right)P_{\mathrm{e}}(\delta_\mathrm{fb}+1),
\end{align}
with $P_{\mathrm{e}}(i)$ as an error probability at a certain \ac{RV}, which is $P_\mathrm{e}(T_\mathrm{max}) := P_{\epsilon_i}$ for $i=T_\mathrm{max}$, and otherwise is defined recursively as:
\begin{align}
    P_{\mathrm{e}}(i) &:=
        P_{\epsilon_i}(P_{\mathrm{fp}_i} + (1-P_{\mathrm{fp}_i})P_{\mathrm{e}}(i+1)).
\end{align}
The stated problem in (\ref{eq:prHARQ_opt}) is a multi-variate optimization problem with non-linear constraints. Since the functional mapping between the false-positive and the false-negative error probabilities is unknown, however can be extracted numerically from the link-level simulations, we used the trust-region constrained algorithm to find local minima to the stated problem with non-linear constraints \cite{trusted_region}. A random starting point has been chosen and the Monte-Carlo method has been used to find a near-optimal solution.

\section{Results}
\label{sec:results}
We performed link-level simulations to evaluate the efficiency of the different \ac{HARQ} schemes. A variety of scenarios with different \acp{SNR}, target \acp{BLER}, \ac{TB} sizes and feedback delays have been modelled and the \acp{EEG} of the different \ac{HARQ} schemes have been determined individually for each scenario. In the first part, we evaluated the \ac{HARQ} approaches under the assumption of the same feedback delay for all \ac{HARQ} schemes. In the second part, we compared the performance of the \ac{prHARQ} scheme to the \ac{paHARQ}, assuming a feedback delay reduction by one time unit $\delta_\mathrm{t}$, i.e. one \ac{OFDM} symbol.

\subsection{Performance evaluation of the HARQ schemes with same feedback delay}
\label{sec:pharq-rharq}
\begin{figure}
  \centering
  \includegraphics[width=0.45\textwidth]%
    {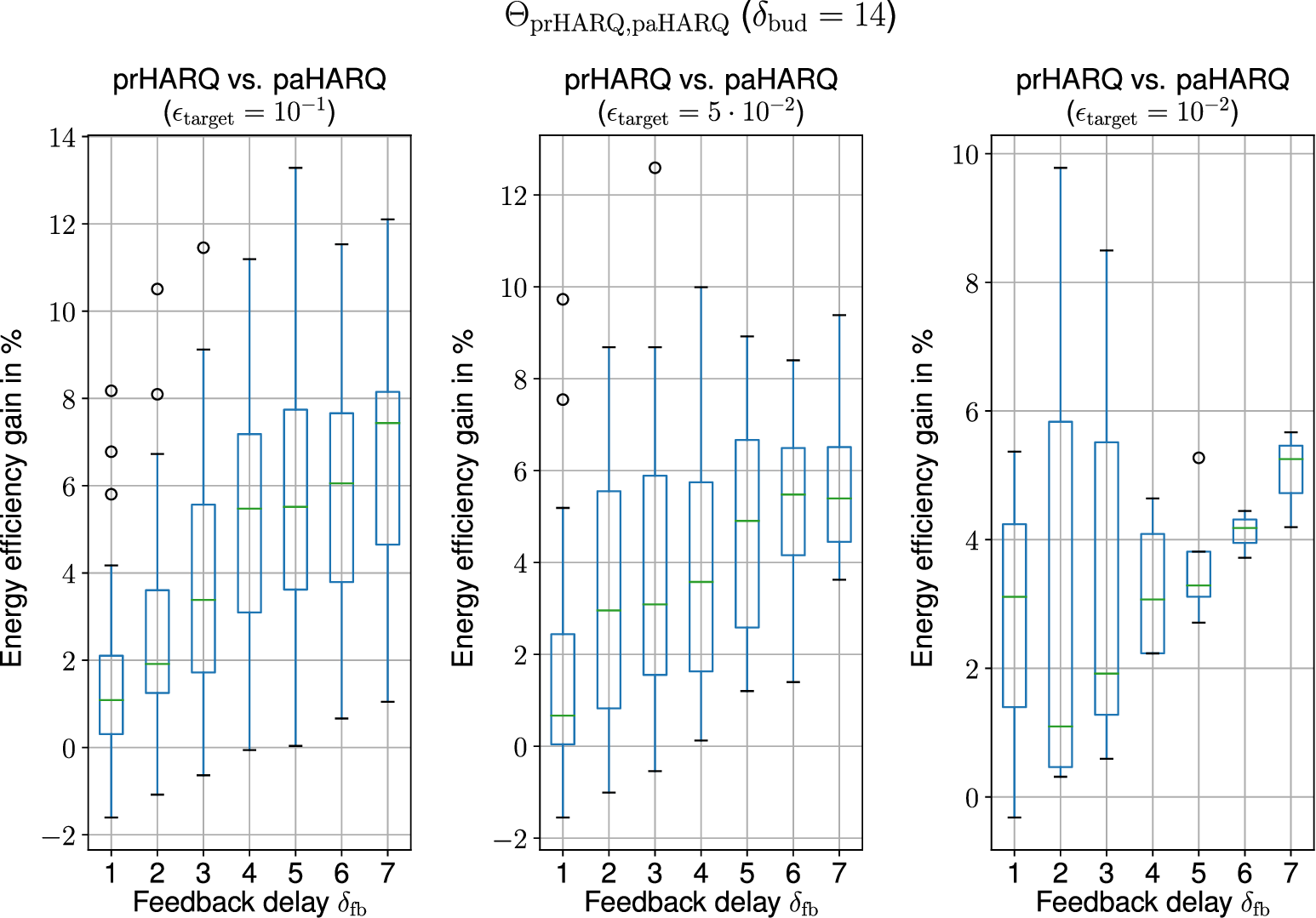}
\caption{\ac{EEG} of the \ac{prHARQ} approach compared to the \ac{paHARQ} over different feedback lengths for all transport block sizes, \acp{SNR}, and $\delta_\mathrm{bud} = 14$.}
\label{fig:th_gain_sa1}
\end{figure}
Fig.~\ref{fig:th_gain_sa1} presents the \ac{EEG} of \ac{prHARQ} compared to \ac{paHARQ} over different feedback delays for three target \acp{BLER}. The \ac{EEG} stays in approximately the same range for all three \ac{BLER} targets. It is clearly visible that the average \ac{EEG} shows a gain for all feedback delays. For feedback delays $\delta_\mathrm{fb}$ larger than three, the \ac{EEG} is solely positive for even all evaluation points. Furthermore, an increase of the gain for higher feedback delays is consistently notable for all the target error rates. Hence, it can be concluded that the prediction provides a significant efficiency gain compared to the non-predictive case even under the assumption of no processing time reduction for the prediction.

\begin{figure}
  \centering
  \includegraphics[width=0.45\textwidth]%
    {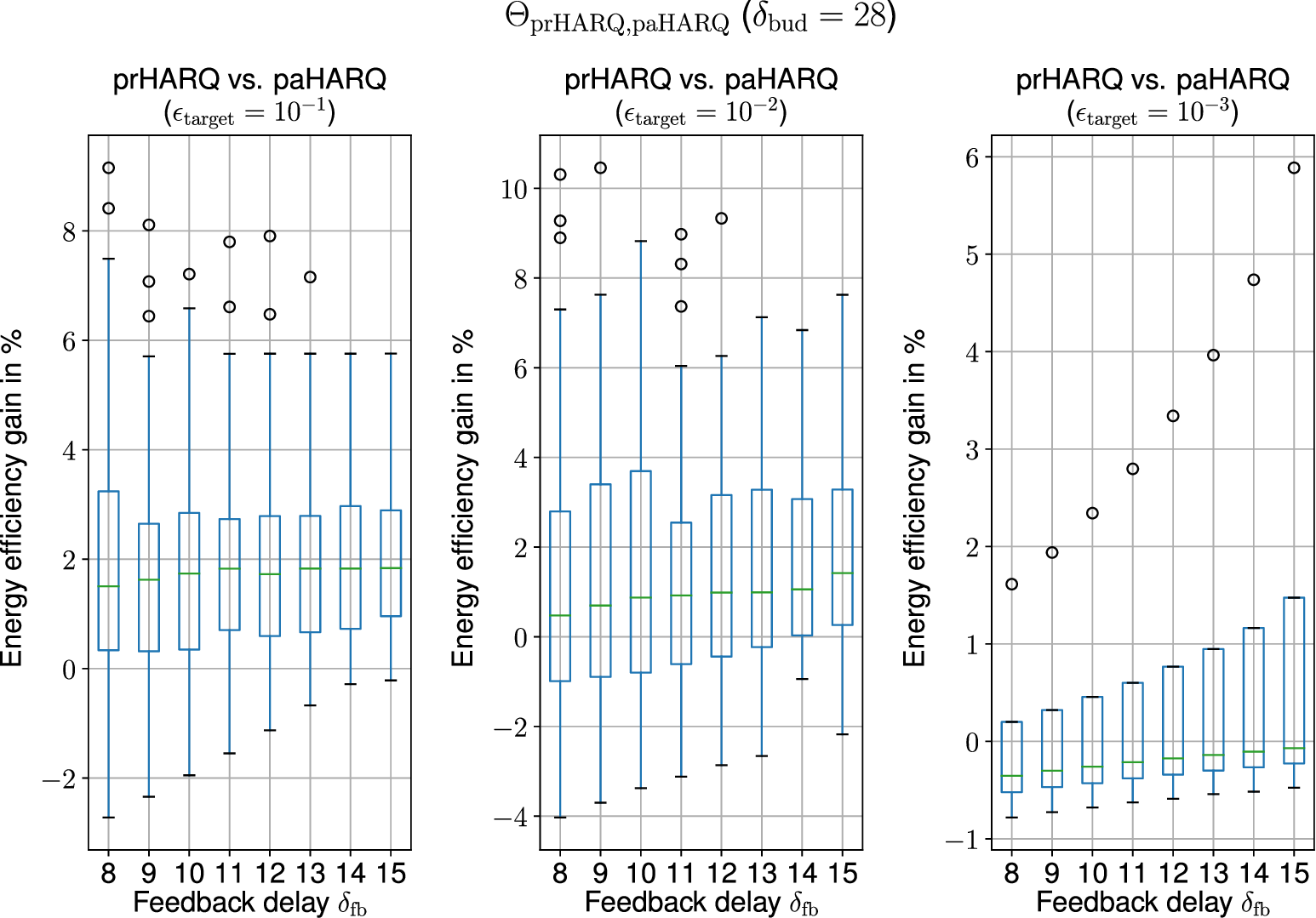}
\caption{\ac{EEG} of the \ac{prHARQ} approach compared to the \ac{paHARQ} over different feedback lengths for all transport block sizes, \acp{SNR}, and $\delta_\mathrm{bud} = 28$.}
\label{fig:th_gain_sa2}
\end{figure}
As can be seen in Fig.~\ref{fig:th_gain_sa2}, the \ac{EEG} is in a similar\added{,} however slightly smaller\added{,} range for the delay budget of 28 OFDM symbols for the different \ac{BLER} targets as compared to the more stringent latency budget. For the target error rate of $\epsilon_\mathrm{target} = 10^{-1}$, the \ac{EEG} is mainly in the range of 0\%-3\% with a slightly increasing trend for higher feedback delays. The \ac{EEG} increases on average for higher feedback delays. For the lowest target error rate of $\epsilon_\mathrm{target} = 10^{-3}$, the \ac{EEG} decreases generally compared to the higher \ac{BLER} regimes. In contrast to the scenario of a very stringent latency target, there are more evaluation points with a negative gain than a positive one even for high feedback delays. This suggests that feedback prediction without processing time reduction compared to full decoding may only be beneficial under certain circumstances, such as a very tight latency target and rather higher \acp{BLER}.
\begin{figure}
  \centering
  \includegraphics[width=0.45\textwidth]%
    {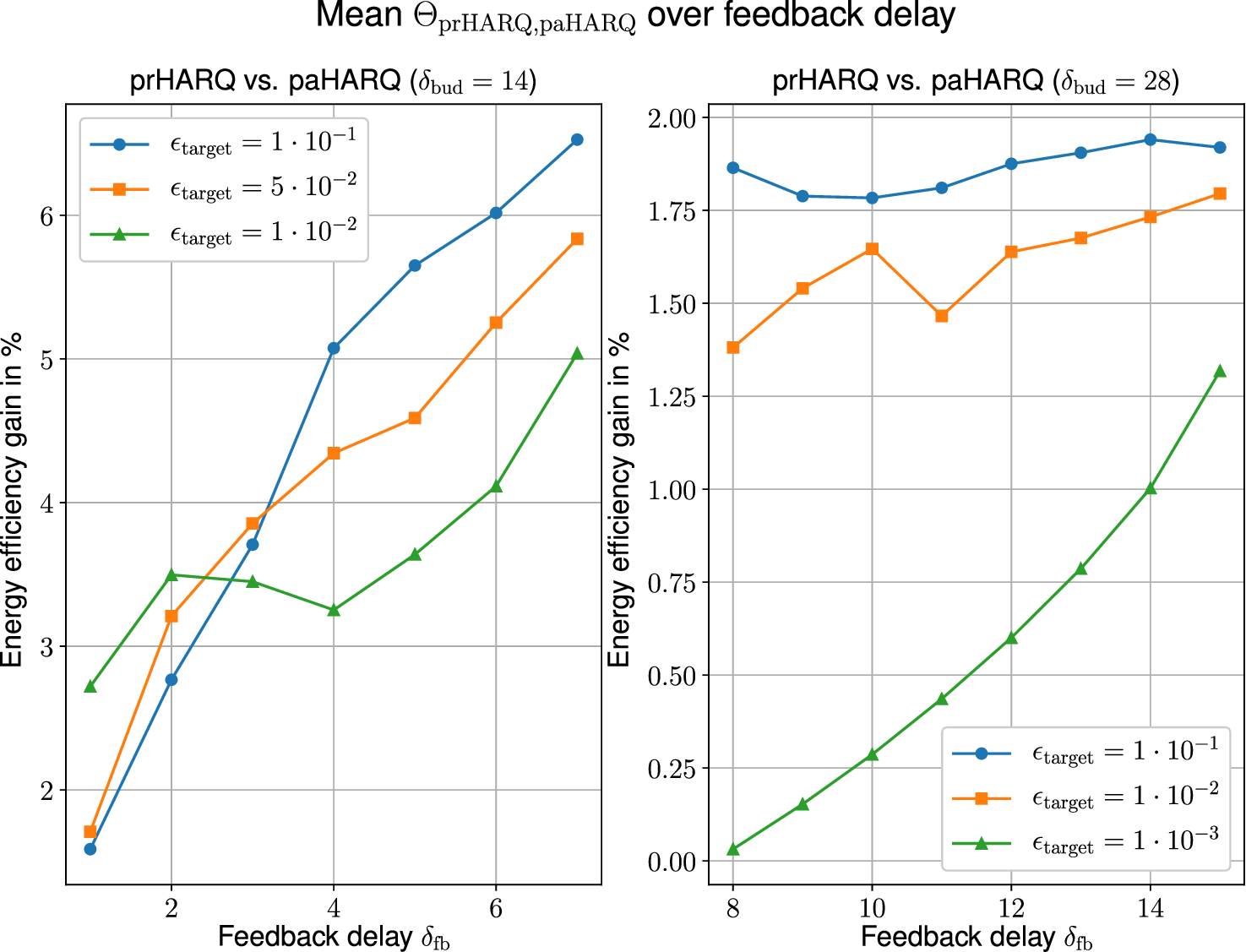}
\caption{Mean \ac{EEG} of the \ac{prHARQ} approach compared to the \ac{paHARQ} over different feedback lengths.}
\label{fig:mean_gain}
\end{figure}
In Fig.~\ref{fig:mean_gain}, the same trends are also observable for the mean \ac{EEG}. While showing solely a positive gain for both latency constraints, the mean \ac{EEG} of the very stringent delay budget is clearly higher compared to less stringent one. In the very stringent scenario, the \ac{prHARQ} approach performs better at lower \ac{BLER} targets for very small feedback delays whereas this trend changes to the contrary at medium and high feedback delays which hints to that the prediction accuracy is decreasing relatively stronger with increasing feedback delay in the lower \ac{BLER} regime.

\subsection{Evaluation of early feedback}
\label{sec:early_feedback}

\begin{figure}
  \centering
  \includegraphics[width=0.45\textwidth]%
    {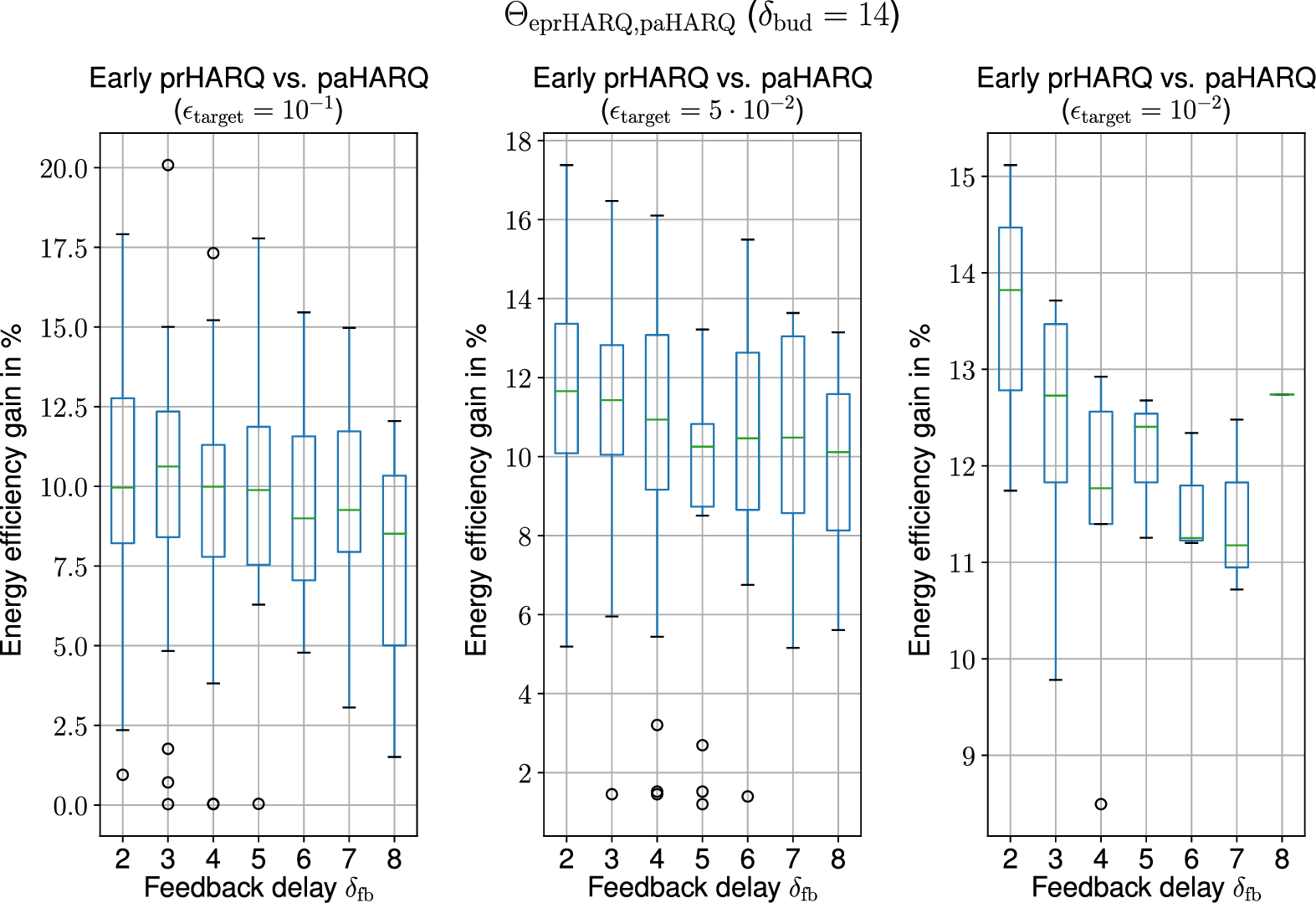}
\caption{\ac{EEG} of the \ac{eprHARQ} approach compared to the \ac{paHARQ} over different feedback lengths for all transport block sizes, \acp{SNR}, and $\delta_\mathrm{bud} = 14$.}
\label{fig:early_sa1}
\end{figure}
One of the major motivations to perform a low-complexity prediction is to reduce the processing time and, thus reduce the feedback delay. In this section, we analyze the gain of the \ac{eprHARQ} scheme assuming a feedback delay reduced by one time unit, i.e. one \ac{OFDM} symbol. In Fig.~\ref{fig:early_sa1} the \ac{EEG} is shown over the feedback delay of the \ac{paHARQ} scheme for $\delta_\mathrm{bud}$=14. As clearly notable, the \ac{eprHARQ} scheme profits significantly from the reduced feedback delay and the \ac{EEG} is for most of the evaluation points in the range of approximately 8\% and 14\%. In general, the lower target \acp{BLER} profit even more from the reduced feedback delay. However in contrast to the case without processing time reduction, higher feedback delays profit less than lower feedback delays. This is explainable by the fact that the reduction of the feedback delay by one \ac{OFDM} symbol is relatively smaller for higher feedback delays. Overall, a clear benefit for the energy efficiency of the \ac{eprHARQ} scheme can be observed.

\begin{figure}
  \centering
  \includegraphics[width=0.45\textwidth]%
    {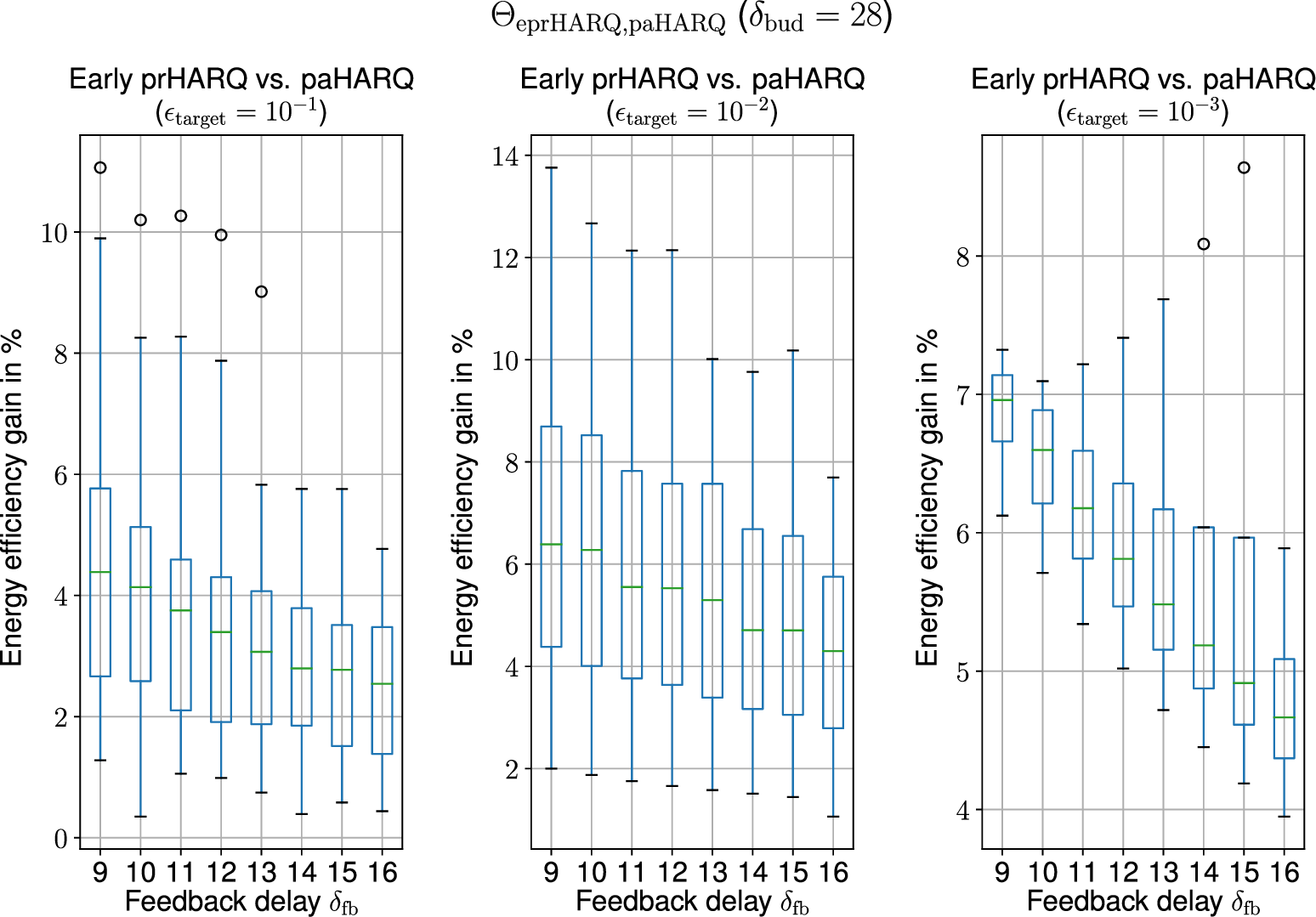}
\caption{\ac{EEG} of the \ac{eprHARQ} approach compared to the \ac{paHARQ} over different feedback lengths for all transport block sizes, \acp{SNR}, and $\delta_\mathrm{bud} = 28$.}
\label{fig:early_sa2}
\end{figure}
The same effects are also observable for the latency budget of 28 \ac{OFDM} symbols in Fig.~\ref{fig:early_sa2}. However, compared to the more stringent latency constraint the \ac{eprHARQ} profits less overall. This can be explained by the relatively smaller reduction of the feedback delay compared to the latency budget. Nevertheless, the \ac{EEG} is in the range of approximately 1\% up to 12\% for most evaluation points. For the lowest target \ac{BLER} of $\epsilon_\mathrm{target} = 10^{-3}$, a significant \ac{EEG} between 4.5\% up to 7.5\% is achieved by the \ac{eprHARQ} scheme, where the gain tends to decrease with an increasing feedback delay. To summarize \ac{eprHARQ} achieves a significant \ac{EEG} being slightly larger for the very stringent latency requirement compared to the less stringent one. This suggests that the strength of \ac{eprHARQ} mainly lies in scenarios with low \ac{BLER} targets and very stringent latency requirements making it a perfect candidate technique for \ac{IIOT} scenarios.
\begin{figure}
  \centering
  \includegraphics[width=0.45\textwidth]%
    {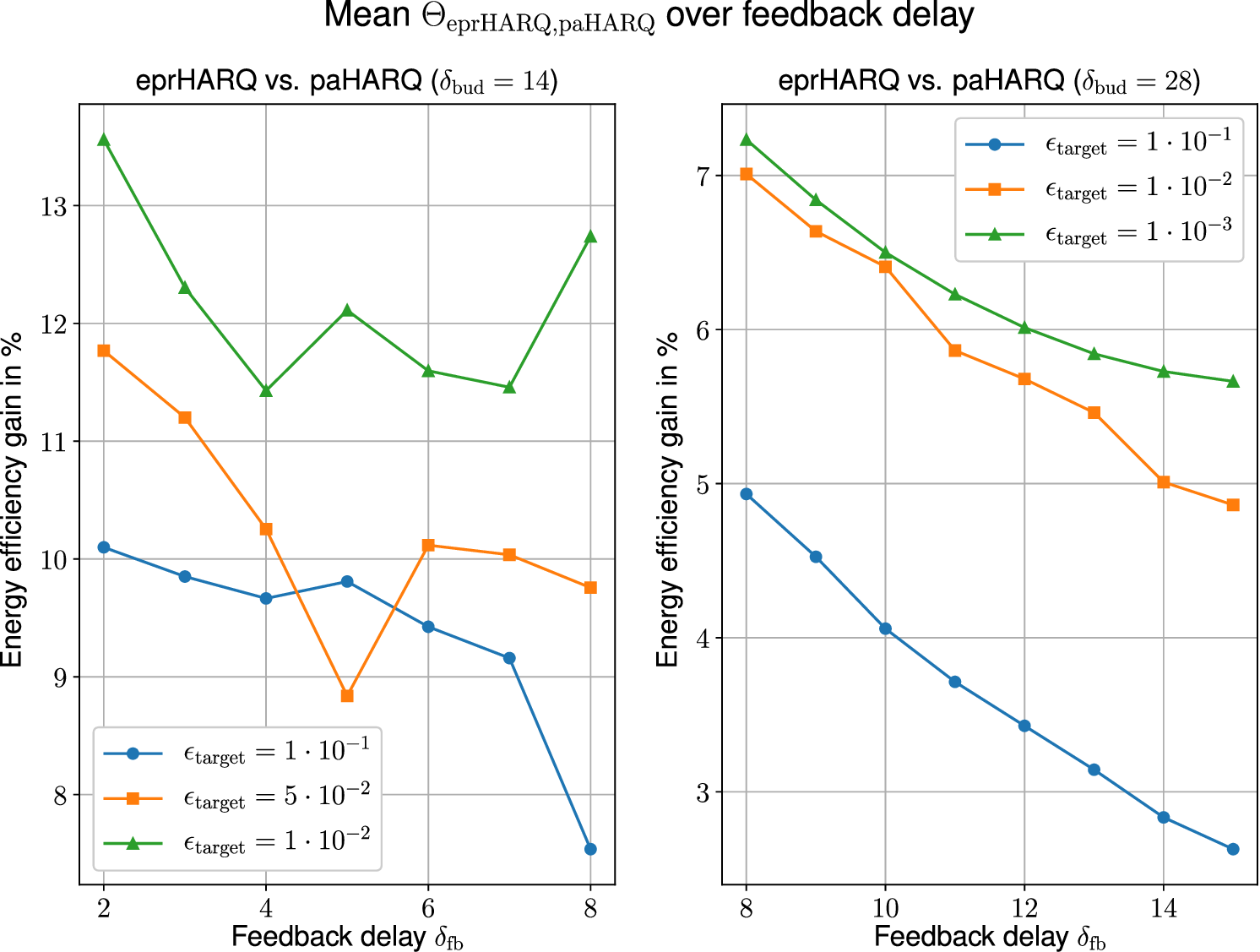}
\caption{Mean \ac{EEG} of the \ac{eprHARQ} approach compared to the \ac{paHARQ} over different feedback lengths.}
\label{fig:mean_gain_early}
\end{figure}
Fig.~\ref{fig:mean_gain_early} shows the mean \ac{EEG} over the different feedback delays. As already noted in the previous figures, the mean \ac{EEG} is decreasing with increasing feedback delay which hints to that the processing time reduction becomes relatively less significant with increasing feedback delay. However, the \ac{eprHARQ} approach achieves a significant mean \ac{EEG} for both latency constraints over all \ac{BLER} targets and all feedback delays, especially in the lower \ac{BLER} regime as noted previously.

\subsection{Comparison of proactive and early proactive HARQ with reactive HARQ}

\begin{figure}
  \centering
  \includegraphics[width=0.45\textwidth]%
    {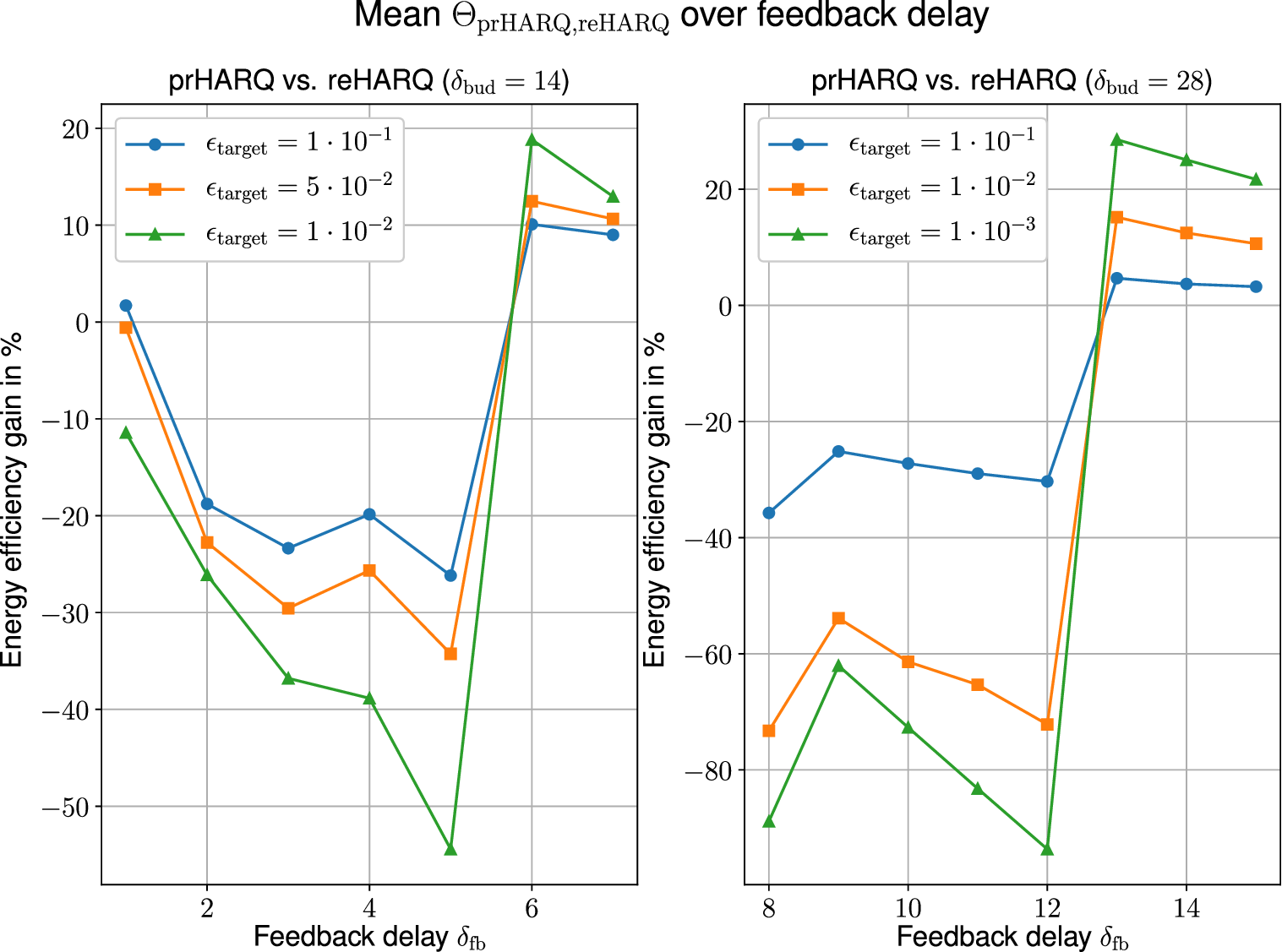}
\caption{Mean \ac{EEG} of the \ac{prHARQ} approach compared to the \ac{reHARQ} without power constraint over different feedback lengths for all transport block sizes, \acp{SNR}, and delay budgets.}
\label{fig:pharq-charq}
\end{figure}
In Fig.~\ref{fig:pharq-charq} the mean \ac{EEG} of \ac{prHARQ} compared to \ac{reHARQ} is shown over the feedback delay. As explained previously, the performance of the \ac{reHARQ} scheme is obtained under the assumption of no power limitations at the transmitter device. Still a clear positive gain of the \ac{prHARQ} approach is notable for sufficiently high feedback delays of 6 and 13 for a latency budget of 14 and 28, respectively. A higher gain for lower target \acp{BLER} is clearly notable, which is due to the low efficiency of the \ac{reHARQ} in this particular case. For smaller feedback delays, \ac{reHARQ} clearly outperforms the \ac{prHARQ} scheme by avoiding unnecessary retransmissions. However, due to power or complexity limitations the \ac{reHARQ} scheme may not be suitable for \ac{IIOT} devices, as discussed above.
\begin{figure}
  \centering
  \includegraphics[width=0.45\textwidth]%
    {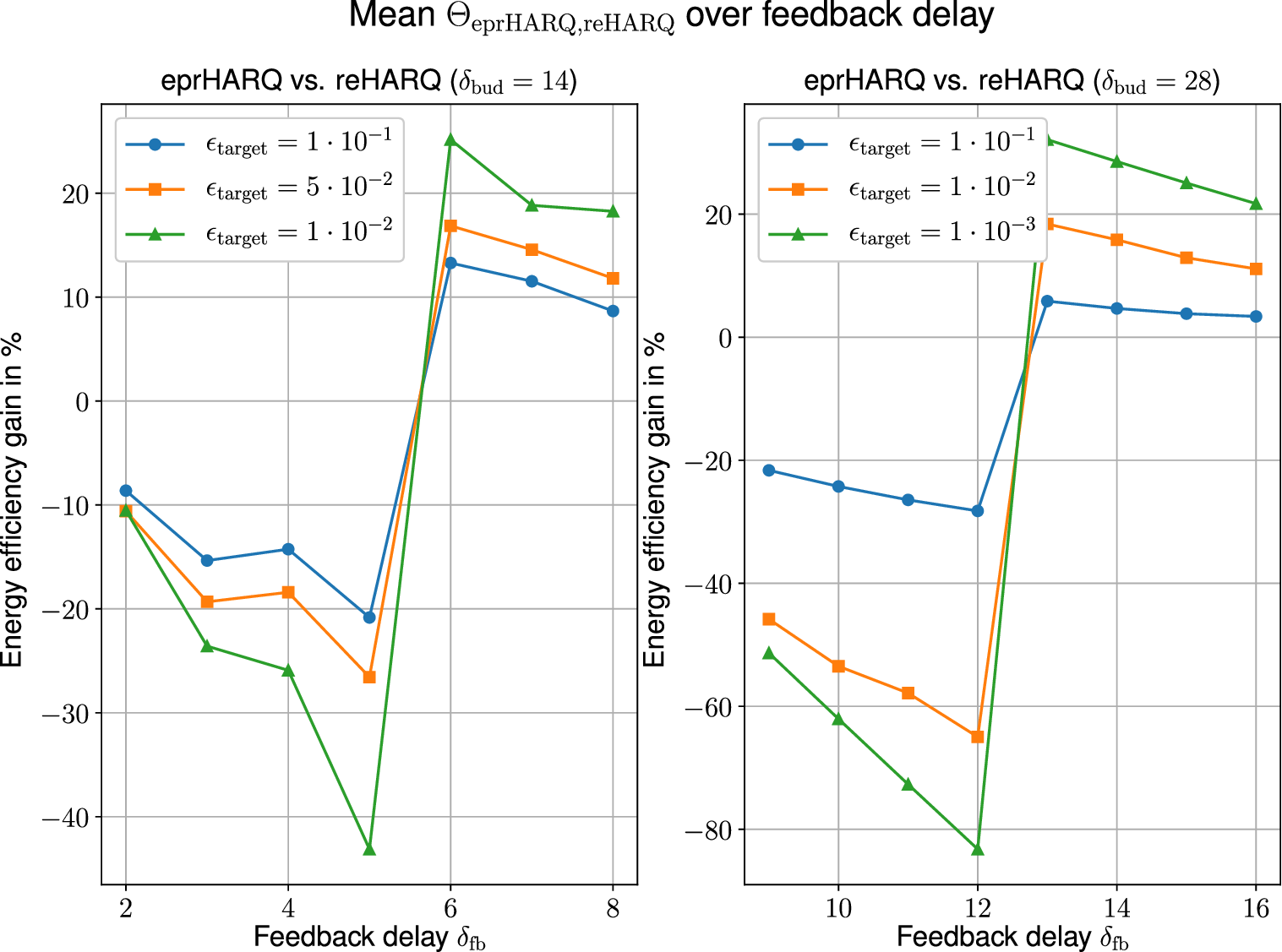}
\caption{Mean \ac{EEG} of the \ac{eprHARQ} approach compared to the \ac{reHARQ} without power constraint over different feedback lengths for all transport block sizes, \acp{SNR}, and delay budgets.}
\label{fig:pharq-charq-early}
\end{figure}
In Fig.~\ref{fig:pharq-charq-early}, we can see for the early feedback case that the \ac{eprHARQ} outperforms the \ac{reHARQ} again at the feedback delays of 6 and 13 for more stringent and less stringent delay budget, respectively. Compared to the proactive scheme without early feedback, it is observable that the efficiency gain increases at the same feedback delays, which is due to the early prediction. Despite to the higher complexity and power requirements, \ac{reHARQ} is outperformed by proactive \ac{HARQ} with or without processing time reduction at sufficiently large feedback delays. 

\section{Conclusion}
We evaluated the energy efficiency of \ac{prHARQ} compared to \ac{paHARQ}. We showed that the prediction improves the energy consumption, demonstrating more advantage for a higher \ac{BLER} and less advantage for a lower \ac{BLER} at least if no processing time reduction is considered. In general, the performance of the \ac{prHARQ} improved for increasing feedback delays hinting at the benefits of the prediction to cope especially with high feedback delays. In the second part, we showed that considering a small feedback delay reduction by one \ac{OFDM} symbol for the feedback prediction mechanism, the \ac{EEG} increases significantly in favor of the \ac{eprHARQ}. Especially in the low \ac{BLER} regime, the energy efficiency is increased by 11\% up to 15\% and 4\% up to 7.5\% for the very stringent and stringent latency constraint, respectively. Hence, also the energy consumption is reduced significantly. In the last part, we demonstrated that even unconstrained \ac{reHARQ} cannot outperform power-constrained \ac{prHARQ} for sufficiently large feedback delays of 6 and 13 \ac{OFDM} symbols for the more stringent and the less stringent latency budget, respectively.

\bibliographystyle{IEEEtran}
\bibliography{lib}

\appendices


\end{document}